\documentclass[aps,
superscriptaddress,
prb,
amsmath,
footinbib,
amssymb,
bibnotes,
twocolumn
]{revtex4-2}

\usepackage{graphicx}
\usepackage{soul}      
\usepackage{color}
\usepackage[colorlinks, citecolor=blue,linkcolor=blue,urlcolor=blue]{hyperref}
\usepackage{verbatim}
\usepackage{multirow}

\graphicspath{{TmpFigs/}}

\usepackage{relsize}
\usepackage{lipsum}
\relscale{1.5}

 \usepackage{mathtools}
 \usepackage{wrapfig} 
 \usepackage{mathrsfs}
 \usepackage{amssymb} 
 \usepackage[x11names]{xcolor}
 \usepackage{amsbsy}
 \usepackage{marginnote}
 \usepackage{slashed}
 \usepackage{marvosym}
 \usepackage{pifont}
 \usepackage{mathbbol}
 \usepackage{wrapfig}
 \usepackage{upgreek}
 \usepackage{bbm}
 \usepackage{yfonts}
 \usepackage{xspace}
 \usepackage{lineno}
 \usepackage{tikz}
\usepackage[most]{tcolorbox}
 \usepackage{siunitx}
 \usepackage{chemformula}

 \newcommand{\bcen}{\begin{center}}
 \newcommand{\ecen}{\end{center}}
 \newcommand{\btab}{\begin{tabular}}
 \newcommand{\etab}{\end{tabular}}
 \newcommand{\bdes}{\begin{description}}
 \newcommand{\edes}{\end{description}}

 \newcommand{\beq}{\begin{equation}}
 \newcommand{\eeq}{\end{equation}}
 \newcommand{\bea}{\begin{eqnarray}}
 \newcommand{\eea}{\end{eqnarray}}

 \newcommand{\bary}{\begin{array}}
 \newcommand{\eary}{\end{array}}
 \newcommand{\benum}{\begin{enumerate}}
 \newcommand{\eenum}{\end{enumerate}}
 \newcommand{\bitem}{\begin{itemize}}
 \newcommand{\eitem}{\end{itemize}}

\let\chapter\section
\let\section\subsection
\let\subsection\subsubsection

\newcommand{\oibook}[1]{}

\newcommand{\term}[1]{\left( #1 \right)}

\newcommand{\eqnref}[1]{Eq.~(\ref{#1})}

\newcommand{\figref}[1]{Fig.~\ref{#1}}
\newcommand{\sfigref}[2]{Fig.~\hyperref[#1]{\ref{#1}#2}}

\newcommand{\Caltech}{Department of Physics and Institute of Quantum Information and Matter, California Institute of Technology, Pasadena, CA 91125, USA.}

\newcommand{\mytitle}{Three-dimensional quantum Hall states as a chiral electromagnetic filter}

\begin{document}

\title{\mytitle}

\author{Nandagopal Manoj}\email{nmanoj@caltech.edu}
\author{Valerio Peri}
\affiliation{\Caltech}

\date{\today{}}    
\begin{abstract} 
Extensive research has explored the optical properties of topological insulating materials, driven by their inherent stability and potential applications. In this study, we unveil a novel functionality of three-dimensional integer quantum Hall (3D IQH) states as broad-band filters for circularly polarized light, particularly effective in the terahertz (THz) frequency range under realistic system parameters. We also investigate the impact of practical imperfections, demonstrating the resilience of this filtering effect.
Our findings reveal that this phenomenon is independent of the microscopic origin of the 3D IQH state, prompting discussions on its feasibility across diverse candidate materials.
These results contribute to our understanding of fundamental optical properties and hold promise for practical applications in optical technologies.
\end{abstract}

\maketitle 



\noindent
\emph{\bf Introduction.} Topological insulating states present promising opportunities for investigating unconventional optical responses \cite{Hasan2010,Wu2016,Qi2011}. Their lack of bulk longitudinal conductivity prevents optical absorption, while the topological properties of the bulk material can profoundly influence the propagation of electromagnetic waves within it.
Early studies on topological matter focused on understanding the impact of a non-zero Chern number on the optical response of 2D systems. These materials exhibit intriguing phenomena such as the Faraday and Kerr effects, where the polarization plane of incident linearly polarized light rotates upon transmission or reflection, respectively \cite{Tse2011}. These effects arise from an additional term in the Lagrangian, namely the Chern-Simons term, which modifies the action of the electromagnetic field at low frequencies \cite{FrohlichKerler1991,Zhang1989,Girvin1987,Zhang1992}.
In 3D strong topological insulators, a modest magnetic field that induces a gap in the surface states leads to Faraday and Kerr effects similar to those observed in 2D Chern insulators \cite{Tse2011, Qi2008, Tse2010}. Here, these effects stem from an axion term, coupling the electric and magnetic fields \cite{Qi2008,Essin2009,Wilczek1987}. While this modification has little impact on the bulk of 3D topological insulators at low energies, it produces observable consequences on their surfaces.
\begin{figure}
    \centering
    \includegraphics{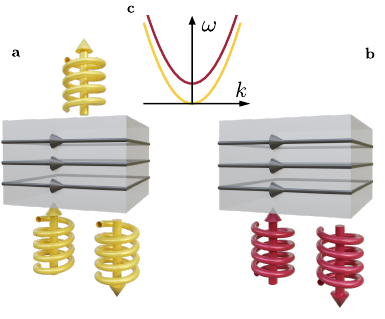}
    \caption{Schematic representation of the chiral electromagnetic filter within a three-dimensional integer quantum Hall state. {\bf a} Right (yellow) circularly polarized light undergoes transmission and reflection. {\bf b} Left (red) circularly polarized light is entirely reflected. The chirality of the transmitted polarization correlates with the chirality of the 3D integer quantum Hall state. {\bf c} Dispersion relation of right (yellow) and left (red) circularly polarized photons. The dispersion relation of the left circular polarization exhibits a gap. \label{fig:1}}
\end{figure}

An intriguing question arises: Is there a possibility of a 3D insulating topological state where the bulk itself substantially influences the propagation of electromagnetic waves? To address this question, we investigate the 3D integer quantum Hall (IQH) effect. This state is characterized by a vanishing bulk longitudinal conductivity and a non-zero in-plane Hall conductivity
\begin{equation}
	\sigma_{xy}=\frac{e^2K}{hd},
	\end{equation}
with $d$ the layer spacing in the out-of-plane direction, and $K$ the layer Chern number \cite{Halperin1987,Balents1996}. The 3D IQH effect was initially predicted as the ground state of an interacting 3D electron gas in the ultra-quantum limit, i.\,e., in a quantizing magnetic field that confines the electrons to the lowest Landau level \cite{Celli1965,Fukuyama1978, Yakovenko1993}. Its first experimental realization, however, came in the form of a stack of 2D IQH states in a semiconductor multilayer superstructure of \ch{GaAs/(AlGa)As} \cite{Stromer1986,Druist1998}. Evidence of 3D IQH behavior was also reported in bulk materials in a strong external field in inorganic Bechgaard salts \cite{Cooper1984}, $\eta$-\ch{Mo4O11} \cite{Hill1998}, n-doped \ch{Bi2Se3} \cite{Cao2012}, \ch{EuMnBi2} \cite{Masuda2016}, and graphite \cite{Yin2019,Fauque2013}. Recently, low-density Dirac semimetals as \ch{ZrTe5} \cite{Tang2019} and \ch{HfTe5} \cite{Galeski2020,Wang2020} emerged as promising candidates to observe the 3D IQH state as they reach the ultra-quantum limit at relatively small external fields, i.\,e., $B=\SI{1.3}{\tesla}$ and $B=\SI{1.8}{\tesla}$, respectively. In all these materials, however, the interpretation of the transport data as a smoking-gun signature of 3D IQH effect remains controversial due to a non-vanishing longitudinal conductivity \cite{Galeski2021,Fang2020,Gooth2023}. 

In this study, we focus on the optical properties of the 3D IQH state. We develop a low-energy field theory for the electromagnetic field in this system, indifferent to its microscopic realization. We demonstrate that the topological properties of the gapped fermions modify the Maxwell action, resulting in a filter effect for circularly polarized light across a wide range of frequencies. Specifically, only light with a defined circular polarization is transmitted, while the opposite polarization is entirely reflected, cf.~\figref{fig:1}. This behavior goes beyond conventional Faraday and Kerr effects, as one circular polarization's dispersion relation becomes gapped within the material's bulk, as shown in \figref{fig:1}~{\bf c}. This filter could serve as an optical circular polarizer in frequency regimes where existing polarizers are inefficient, and it could aid in identifying 3D IQH states in challenging-to-contact materials where transport studies alone may be inconclusive. 

In the following, we describe our field-theoretic model for a 3D IQH state as a stack of 2D IQH layers coupled to the electromagnetic field. We start by reviewing the study of light-matter interactions in a single 2D Chern insulator, as it serves as the foundational element for our subsequent analysis. After presenting the main result of our work, we examine two non-ideal factors that can influence the efficiency of the proposed filtering mechanism: variations in the incidence angle and finite longitudinal conductivity. Finally, we summarize our findings and discuss the prospects of experimentally observing the predicted filtering effect. 

\noindent
\emph{\bf Model.} To describe the interaction of a 2D Chern insulator with Chern number $K$ with the electromagnetic field at energies below the charge gap, we integrate out the gapped fermions and get an effective action for the electromagnetic field alone \cite{Zhang1989,FrohlichKerler1991}. The topological bands contribute a Chern-Simons term with the coefficient $K$ to the effective action
\begin{multline}
    S[A] = \int d^3xdt \Bigg( \sum_{\mu,\nu = 0,1,2,3}\frac{1}{16\pi} F_{\mu\nu}F^{\mu\nu}  \\
    -  \sum_{\mu,\nu,\lambda = 0,1,2}\frac{\alpha K\delta(x_3)}{4\pi} \epsilon^{\mu \nu \lambda} A_\mu  \partial_\nu A_\lambda \Bigg)\,,
\end{multline}
where $F_{\mu\nu} = \partial_\mu A_\nu - \partial_\nu A_\mu $ is the electromagnetic field tensor and  $\alpha\approx 1/137$ is the fine structure constant. 
Solving the equations of motion for light propagating along $\hat{z}$, i.\,e., the out-of-plane direction, with angular frequency $\omega$ gives the complex transmission and reflection amplitudes
\begin{equation}
\begin{split}
    t_\pm &=\frac{1 \pm i\alpha K }{1 + \alpha^2  K^2}\,, \\
    r_\pm &= \pm  \frac{i \alpha K\left(1 \pm i\alpha K\right)}{1 + \alpha^2  K^2}\,,
    \end{split}
\end{equation}
for right circularly polarized (RCP, $+$) and left circularly polarized (LCP, $-$) light, respectively. 
The two polarizations do not mix because the rotational symmetry about the propagating axis is preserved. 
Since the fine structure constant $\alpha $ is small, the effect of a single layer is quite weak and most of the light is transmitted. 
Moreover, the magnitude of the reflection $R_\pm=\lvert r_\pm\rvert^2$ and transmission $T_\pm=\lvert t_\pm\rvert^2$ coefficients are frequency-independent and equal for both polarizations. 
Crucially, both the transmission and reflection amplitudes for the two modes are related via complex conjugation and the two polarizations pick up opposite phases. These phases indicate a rotation of the polarization vector along the axis of propagation -- a rotation of the plane of polarization for linearly polarized light. This is the origin of the Faraday and Kerr effects, which predict a rotation angle
\begin{equation}
    \theta_{\text{F}} = \text{arctan}(\alpha K), \quad \theta_{\text{K}} = \mathrm{sgn}(K)\frac{\pi}{2} +  \text{arctan}(\alpha K),
\end{equation}
for transmission and reflection, respectively \cite{Tse2011}.

To explore whether there exists a setting in which these phases can also selectively impact the reflection ($R$) and transmission ($T$) coefficients, we turn to a stack of $N$ Chern layers with layer spacing $d$. This system is described by the action
\begin{multline}
    S[A] = \int d^3xdt \Bigg( \sum_{\mu,\nu = 0,1,2,3}\frac{1}{16\pi} F_{\mu\nu}F^{\mu\nu} \\
    - \sum_{j = 0}^{N-1} \sum_{\mu,\nu,\lambda = 0,1,2}\frac{\alpha K\delta(x_3 - jd)}{4\pi} \epsilon^{\mu \nu \lambda} A_\mu  \partial_\nu A_\lambda \Bigg).
\end{multline}
The two polarization modes pick up the same phase $e^{-ikd}$ while propagating between two layers. 
We anticipate that this phase, along with the opposite phases induced by the individual Chern layer, alters the interference of the two polarization modes when multiple layers are present. This interference effect becomes significant in the limit of many layers and leads to the filtering effect. 
\begin{figure}
    \centering
    \includegraphics{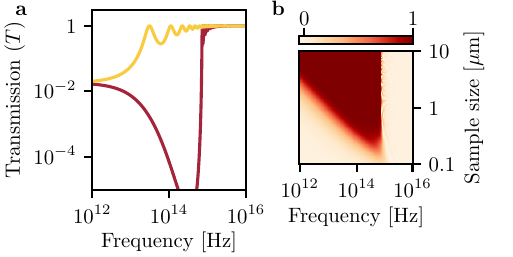}
    \caption{Chiral electromagnetic filter effect demonstrated in a slab of 3D integer quantum Hall state with a layer Chern number $K=1$ and a layer separation of $d = \SI{1}{\nano\meter}$. {\bf a} Transmission coefficient for right circularly polarized (RCP, yellow) and left circularly polarized (LCP, red) light with $N = 1024$ layers. {\bf b} Transmission contrast $(T_+-T_-)/(T_++T_-)$ as a function of frequency and sample thickness.
    \label{fig:2}}
\end{figure}

To calculate the effective transmission resulting from the interference, we numerically sum the probability amplitudes of all the possible paths that the photon can take. Figure~\ref{fig:2} {\bf a} displays the result of this calculation. Notably, there is a strong filtering effect over an optimal frequency range: the transmission of the LCP mode is suppressed by many orders of magnitude. One can choose the polarization of the filtered mode by switching the chirality of the Chern layers. 

The above result suggests that the relevant physics takes place at light wavelengths much larger than the layer separation. We then expect the optical response to be independent of the microscopic origin of the Chern layers. We are therefore justified in smearing out the individual Chern-Simons terms to obtain a bulk effective theory for describing the long-wavelength photons in the 3D IQH state:
\begin{multline}
	\label{eq:action3D}
    S[A] = \int d^3xdt \Bigg( \sum_{\mu,\nu = 0,1,2,3}\frac{1}{16\pi} F_{\mu\nu}F^{\mu\nu} \\ - \sum_{\mu,\nu,\lambda = 0,1,2}\frac{\alpha K}{4\pi d} \epsilon^{\mu \nu \lambda} A_\mu  \partial_\nu A_\lambda \Bigg).
\end{multline}
Upon integration by parts, we see that the second term realizes a static axion field ($\theta$-term) \cite{Wilczek1987} that is linearly increasing perpendicular to the layer plane, i.\,e., $\alpha K z/(4\pi d)\boldsymbol{E}\cdot\boldsymbol{B}$. This action is also known as Carroll-Field-Jackiw electrodynamics in the high-energy literature \cite{Carroll1990,Colladay1998}. If the layering spontaneously breaks translation symmetry along $z$, the corresponding Goldstone boson can be interpreted as a dynamical axion field~\cite{Wang2013}.  Note that a similar action describes the low-energy physics of Weyl semimetals, where the linearly growing axion term stems from the momentum-space separation of the Weyl points \cite{Grushin2012,Zyuzin2012,Babak2022}. Indeed, optical effects akin to those presented here were recently discussed in Weyl semimetals \cite{Cote2023,Chtchelkatchev2021}. While the phenomenology is similar, we stress that the microscopic origin of the linearly-growing axion term is markedly different. In particular, the 3D IQH state has a charge gap with a well-defined frequency cut-off below which optical absorption is suppressed.  

From the effective action of \eqnref{eq:action3D}, we can readily calculate the dispersion relation of the photon. For simplicity, we consider the case where the photon is propagating in the $z$ direction, i.\,e. the wavevector $\boldsymbol{k} = k \hat{\boldsymbol{z}}$. Upon solving the equations of motion, we find
\begin{equation}
    c^2 k^2 = \omega^2 \pm \frac{2 \alpha K c}{d}\omega\,,
    \label{eqn:dispersion_relation}
\end{equation}
which has been plotted in \figref{fig:1}~{\bf c}. We note that at low energies, the RCP mode has gapless quadratically dispersing modes whereas the LCP mode acquires a gap $\Delta = 2 \alpha K c/ d$. This result supplies a nice interpretation of the filtering effect. Namely, below the gap frequency, the LCP photon does not have enough energy to enter the bulk and gets reflected, whereas the RCP light can enter and has nontrivial transmission below the gap. At frequencies much below the gap, the LCP light has a penetration depth $\sim \sqrt{dc/(2\alpha K \omega )}$ which means that if the sample is too thin, a significant fraction of LCP photons will transmit across via quantum tunneling, destroying the filter effect. This observation sets a lower bound for sample thickness for effective filtering as shown in \figref{fig:2}~{\bf b}, which displays the transmission contrast $(T_+-T_-)/(T_++T_-)$ between the two polarization modes as a function of frequency and sample thickness. 

Based on the previous discussion, we consider a slab of thickness $Nd$ of the low-energy bulk theory and look at the effective transmission and reflection coefficient across this slab. This is a problem of quantum tunneling across a rectangular barrier. The effective transmission amplitude is
\begin{equation}
    t_{\text{eff}} = \frac{4 k_\text{b} k_\text{v} e^{i Nd (k_\text{b} - k_\text{v})}}{(k_\text{b} + k_\text{v})^2 - e^{ 2 i N d k_\text{b}}(k_\text{b} - k_\text{v})^2}\,,
    \label{eqn:teff1}
\end{equation}
where $k_\text{v} = \omega / c$ is the vacuum wavevector and $k_\text{b}$ is the bulk wavevector given by \eqnref{eqn:dispersion_relation}, both pointing in the $z$ direction. The effective transmission coefficients $T$ for the two modes at low energies are in exact agreement with the numerical calculation in \figref{fig:2}~{\bf a}.

\begin{figure}
    \centering
    \includegraphics{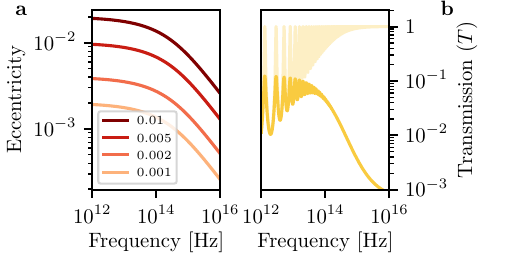}
    \caption{{\bf a} Eccentricity of the transmitted polarization as a function of frequency. The legend reports the values of the angle away from normal incidence in radians. {\bf b} Comparison of transmission coefficients $T$ for the RCP light. The faint curve shows $T$ with $\sigma_{xx}=0$, whereas the bold line represents $T$ with $\sigma_{xx}=\sigma_{xy}/20$. We considered sample thickness $L=\SI{10}{\micro\meter}$ and layer spacing $d=\SI{1}{\nano\meter}$. For the whole frequency range considered, the transmission coefficient of the LCP light is smaller than $10^{-5}$. \label{fig:3}}
\end{figure}

\noindent
\emph{\bf Imperfections. }Having demonstrated the chiral filter effect in the 3D IQH state, we now consider imperfections that might alter our results. Since the separation of modes into LCP and RCP light relied on the rotation symmetry of the problem, the chiral filtering effect is modified when the incident angle is tilted away from normal incidence. We find that while the chirality is retained, the filtered mode becomes elliptically polarized with non-zero eccentricity. 
We solve for the bulk dispersion numerically for propagation at a small angle $\varphi_b$ away from the $z$ axis and find that the propagating photon mode with polarization
\begin{equation}
    \boldsymbol{p}\sim \term{1,\frac{i\omega \Delta  }{\omega^2-c^2k_{\text{b},z}^2\left(1+\varphi_b^2\right)},\frac{-\varphi_b  c^2 k_{\text{b},z}^2}{w^2-\varphi_b^2c^2k_{\text{b},z}^2}}
\end{equation}
becomes linearly dispersing at very low energies with velocity $\varphi_b c$. The orthogonal polarization mode remains gapped. To find the ellipticity of the transmitted polarization, we solve the boundary conditions to calculate the vacuum polarization vector corresponding to the gapless mode in the bulk
\begin{equation}
    \boldsymbol{a} = -\frac{\boldsymbol{k}_\text{v} \times \left[(\boldsymbol{k}_\text{v} + \boldsymbol{k}_\text{b})\times \boldsymbol{p}\right]}{2 \lVert{\boldsymbol{k}_\text{v}}\rVert^2}.
\end{equation}
We plot the eccentricity of the complex vector $\boldsymbol{a}$ by calculating $\boldsymbol{k}_\text{b}$ numerically for a given $\varphi_\text{b}$, and solving Snell's law $\varphi_b = \varphi \lVert \boldsymbol{k}_\text{v} \rVert / \lVert \boldsymbol{k}_\text{b}\rVert$ self-consistently, where $\varphi$ is the tunable parameter, i.\,e., the incidence angle. The result of this analysis is displayed in \figref{fig:3}~{\bf a} for a range of different $\varphi$.

In an ideal 3D IQH state, one expects a vanishing longitudinal conductivity $\sigma_{xx}$ below the charge gap, similar to the two-dimensional case \cite{Halperin1987}. In candidate materials as \ch{ZrTe5} and \ch{HfTe5}, a non-zero longitudinal conductivity is measured even at zero frequency, albeit smaller than the dominant $\sigma_{xy}$ by an order of magnitude. Its origin is still the object of an ongoing debate \cite{Gooth2023,Galeski2021,Qin2020}. For our purposes, a finite $\sigma_{xx}$ causes the absorption of photons passing through the bulk. As the fraction of transmitted photons decays exponentially in sample thickness, it sets an upper bound on the sample's size to observe the filter effect. 

Not knowing the specific origin of the finite $\sigma_{xx}$, we make the simplifying assumption that the conductivity is frequency-independent in the range of interest. Its effect is taken into account by modifying the relative dielectric tensor 
\begin{equation}
    \varepsilon^{(xx)}_r(\omega) = 1 + \frac{i \sigma_{xx}}{\varepsilon_0 \omega}\,,
\end{equation}
which gives us the modified dispersion relation
\begin{equation}
    c^2 k^2 = \omega^2 \pm \frac{2 \alpha K c}{d}\omega + \frac{i \sigma_{xx}}{\varepsilon_0}\omega\,.
    \label{eqn:dispersion_relation2}
\end{equation}
Now, both polarizations have a non-zero imaginary wavevector, indicating a finite penetration depth. This is due to a different mechanism in addition to the previous discussion, namely absorption, which creates exponential suppression of tunneling probability in sample thickness for both polarization modes. One can use \eqnref{eqn:teff1} to once again calculate the transmission coefficient as a function of frequency. Figure~\ref{fig:3} {\bf b} shows the comparison between the transmission coefficient of the RCP light with and without longitudinal conductivity. We considered $\sigma_{xx}=\sigma_{xy}/20$, motivated by the DC transport results of Ref.~\onlinecite{Tang2019}.


\noindent
\emph{\bf Discussions and Conclusions.} To summarize, we found that the 3D IQH state serves as a chiral electromagnetic filter, selectively transmitting only one chirality of circularly polarized light. We elucidated how this effect stems from a linearly growing axion term in the material's bulk, which induces a gap in the dispersion of the photon of the fully reflected circular polarization. This phenomenon distinguishes itself from Faraday and Kerr rotations, which transmit photons with the same probability regardless of polarization.

Understanding the operational frequency range of this filter is crucial. The sample size along the light propagation direction, denoted as $L$, sets a lower angular frequency bound: $\omega \gtrsim dc/(2\alpha K L^2)$. Below such a frequency, quantum tunneling becomes relevant, and both polarizations are transmitted, cf.~\figref{fig:2}. The light then perceives the sample as two-dimensional with an extensively large Chern invariant $K L /d$ and undergoes Kerr and Faraday rotations. Meanwhile, the photon gap determines a cutoff at high angular frequencies, which is inversely proportional to the spacing $a$ between the stacked 2D IQH layers: $\omega < 2\alpha Kc/d$.

In semiconductor multilayer superstructures realizing the 3D IQH effect, $d \approx \SI{22.6}{\nano\meter}$ and $L \approx \SI{0.7}{\micro\meter}$ \cite{Stromer1986}. The sample's thickness successfully blocks one circular polarization only for frequencies above \SI{1e14}{\hertz}. The photon gap, however, sets an upper-frequency limit of \SI{3e13}{\hertz}. Therefore, these platforms do not display the proposed filtering effect at any frequency unless the spacing among the layers is significantly reduced or the thickness of the heterostructure is increased.

Conversely, bulk crystals of \ch{ZrTe5} and \ch{HfTe5} have been grown with sizes up to $L = \SI{300}{\micro\meter}$ \cite{Tang2019,Fang2020,Galeski2020,Galeski2021}. The 3D IQH state in these materials would emerge from a charge-density wave transition where the material spontaneously layers along a specific direction, e.g., $z$. The wavelength of this charge modulation sets the layer spacing and is determined by the Fermi wavevector along the $z$ direction. In these materials, experimental data suggest $d \approx \SI{1}{\nano\meter}$ \cite{Fang2020}. With these parameters, the operational frequency range is $\SI{4e7}{\hertz} \lesssim \nu=\omega/(2\pi) \lesssim \SI{7e14}{\hertz}$. Hence, self-layering materials, where the 3D IQH state emerges as a many-body instability, appear as the most promising candidates to detect the filter discussed in this work. The translation-symmetry breaking is accompanied by a dynamical axion (Goldstone), which we expect to be pinned by disorder. It would be interesting to probe the consequences of this axion-photon coupling in experiments~\cite{Li2010,Nenno2020}.

A more stringent high-frequency limit is set by the charge gap of the system. Specifically, the filter differs from other chiral optical phenomena, such as circular dichroism \cite{Hosur2015}, as it operates at frequencies below any absorption threshold. The presence of a charge gap further distinguishes our proposal from similar ones in Weyl semimetals \cite{Cote2023,Chtchelkatchev2021}. The relevant gap depends on the microscopic realization of the 3D IQH state. For example, in  the charge density wave realization of the 3D IQH effect, both the spacing of the Landau levels and the gap of the amplitude mode of the charge density wave transition are important. The Landau level separation in \ch{ZrTe5} appears to be around $\SI{25}{\milli\electronvolt}$ \cite{Tang2019}, while theory predicts a CDW gap of several tens of \si{\milli\electronvolt} \cite{Qin2020}. Energetic gaps of order of \si{\milli\electronvolt} lead to an upper-frequency limit in the \si{\tera\hertz} range. To then observe a filter effect in this frequency range, samples of size $L>\SI{2}{\micro\meter}$ are necessary.

It is interesting to consider the response of fractionalized phases such as 3D fractional quantum Hall (FQH) states~\cite{NaudPyadkoSondhi2000,LevinFisher2009}. As the smeared out effective action for the bulk is equivalent to that for the 3D IQH state, it does not seem possible that low-energy optical probes can distinguish these more exotic states from 3D IQH states. We leave this as an open problem.

As mentioned, low-density semimetals like \ch{ZrTe5} and \ch{HfTe5} exhibit a finite longitudinal conductivity which sets them apart from an ideal 3D IQH state \cite{Galeski2021}. Any level of longitudinal conductivity restricts the filter's effectiveness by diminishing the transmitted light's amplitude. Nonetheless, our analysis indicates that even with a longitudinal conductivity comparable to experimental observations, the filter effect remains detectable. In such scenarios, it is crucial to select a sample size that prevents excessive attenuation while still ensuring that the material's thickness remains sufficiently larger than the wavelength of the incident light.

In summary, our investigation provides insights into the unique optical characteristics of the 3D integer quantum Hall effect state, unveiling its potential as a chiral electromagnetic filter. Beyond its fundamental significance, this research opens up promising avenues for practical applications in optical technologies. Future investigations may explore alternative approaches to realize the 3D IQH state, potentially independent of magnetic fields. For instance, stacks of 2D anomalous Hall effect or axion charge density wave transitions in Weyl semimetals \cite{Nenno2020,Wang2013,Gooth2019} could be investigated to optimize the performance and applicability of these filters.

\acknowledgments
We acknowledge insightful discussions with J. Alicea, C. B. Dag, A. Grushin, S. D. Huber, P. A. Lee, and P. Moll.
This work was primarily led and supported by the U.S. Department of Energy, Office of Science, National Quantum Information Science Research Centers, Quantum Science Center (N.M.).
V.P. is grateful for the generous support from the Gordon and Betty Moore Foundation’s EPiQS Initiative, Grant GBMF8682. N.M. and V.P. appreciate the support received from the Institute of Quantum Information and Matter.

\bibliographystyle{apsrev4-2}
\bibliography{apssamp}
\end{document}